\documentclass[twocolumn,showpacs,nofootinbib,preprintnumbers,amsmath,amssymb,superscriptaddress,aps,prd]{revtex4-1}

\usepackage{graphicx}
\usepackage{dcolumn}
\usepackage{bm}

\def\lsim{\mathrel{\raise.3ex\hbox{$<$\kern-.75em\lower1ex\hbox{$qf$}}}}
\def\gsim{\mathrel{\raise.3ex\hbox{$>$\kern-.75em\lower1ex\hbox{$\sim$}}}}

\def\Pbeam     {12~\mathrm{GeV/c}}                          
\def\eff       {0.958 \pm 0.011}                   
\def\Nevt      {350}                               
\def\Tcirte    {(16.8 ^{+0.8} _{-1.1})~\mathrm{keV}}  
\def\Tseitz    {(13.6               \pm 0.6)~\mathrm{keV} }       

\begin{document}


\title{Direct Measurement of the Bubble Nucleation Energy Threshold in a 
CF$_3$I Bubble Chamber}

\author{E.~Behnke}
\affiliation{Indiana University South Bend, South Bend, Indiana 46634, USA}
\author{T.~Benjamin}
\affiliation{Indiana University South Bend, South Bend, Indiana 46634, USA}
\author{S.J.~Brice}
\affiliation{Fermi National Accelerator Laboratory, Batavia, Illinois 60510, USA}
\author{D.~Broemmelsiek}
\affiliation{Fermi National Accelerator Laboratory, Batavia, Illinois 60510, USA}
\author{J.I.~Collar}
\affiliation{Enrico Fermi Institute, KICP and Department of Physics,
             University of Chicago, Chicago, Illinois 60637, USA}
\author{P.S.~Cooper}
\affiliation{Fermi National Accelerator Laboratory, Batavia, Illinois 60510, USA}
\author{M.~Crisler}
\affiliation{Fermi National Accelerator Laboratory, Batavia, Illinois 60510, USA}
\author{C.E.~Dahl}
\affiliation{Department of Physics and Astronomy, Northwestern University, Evanston, Illinois 60208, USA}
\affiliation{Fermi National Accelerator Laboratory, Batavia, Illinois 60510, USA}
\author{D.~Fustin}
\affiliation{Enrico Fermi Institute, KICP and Department of Physics,
             University of Chicago, Chicago, Illinois 60637, USA}
\author{J.~Hall}
\affiliation{Pacific Northwest National Laboratory, Richland, Washington 99352, USA.}
\author{C.~Harnish}
\affiliation{Indiana University South Bend, South Bend, Indiana 46634, USA}
\author{I.~Levine}
\affiliation{Indiana University South Bend, South Bend, Indiana 46634, USA}
\author{W.H.~Lippincott}
\email{hugh@fnal.gov}
\affiliation{Fermi National Accelerator Laboratory, Batavia, Illinois 60510, USA}
\author{T.~Moan}
\affiliation{Indiana University South Bend, South Bend, Indiana 46634, USA}
\author{T.~Nania}
\affiliation{Indiana University South Bend, South Bend, Indiana 46634, USA}
\author{R.~Neilson}
\email{rneilson@uchicago.edu}
\affiliation{Enrico Fermi Institute, KICP and Department of Physics,
             University of Chicago, Chicago, Illinois 60637, USA}
\author{E.~Ramberg}
\affiliation{Fermi National Accelerator Laboratory, Batavia, Illinois 60510, USA}
\author{A.E.~Robinson}
\affiliation{Enrico Fermi Institute, KICP and Department of Physics,
             University of Chicago, Chicago, Illinois 60637, USA}
\author{M.~Ruschman}
\affiliation{Fermi National Accelerator Laboratory, Batavia, Illinois 60510, USA}
\author{A.~Sonnenschein}
\affiliation{Fermi National Accelerator Laboratory, Batavia, Illinois 60510, USA}
\author{E.~V\'azquez-J\'auregui}
\affiliation{SNOLAB, Lively, Ontario, Canada P3Y 1N2}
\collaboration{COUPP Collaboration}
\noaffiliation

\author{R.A.~Rivera}
\affiliation{Fermi National Accelerator Laboratory, Batavia, Illinois 60510, USA}
\author{L.~Uplegger}
\affiliation{Fermi National Accelerator Laboratory, Batavia, Illinois 60510, USA}

\date{\today}


\begin{abstract}
We have directly measured the energy threshold and efficiency for bubble 
nucleation from iodine recoils in a CF$_3$I bubble chamber in the energy range of interest for a dark matter search. These interactions cannot be probed by standard neutron calibration methods, so we develop a new technique by 
observing the elastic scattering of $\Pbeam$ negative pions.  The pions are tracked with a silicon pixel telescope and the reconstructed 
scattering angle provides a measure of the nuclear recoil kinetic energy.  The bubble chamber was 
operated with a 
nominal threshold of $\Tseitz$. Interpretation of the results depends on the response to fluorine and carbon recoils, but in general we find agreement with the predictions of the classical bubble nucleation theory. This measurement confirms the applicability of CF$_3$I as a target for spin-independent dark matter interactions and represents a novel technique for calibration of superheated fluid detectors. 
\end{abstract}

\pacs{29.40.-n, 95.35.+d, 95.30.Cq,   FERMILAB-PUB-10-318-A-CD-E}
\maketitle


Recent years have seen a resurgence in the use of superheated liquids and bubble chambers as continuously sensitive 
nuclear recoil detectors searching for dark matter in the form of Weakly Interacting Massive Particles (WIMPs)\cite{COUPPscience, picasso, SIMPLE}. At a low degree of superheat, bubble chambers are insensitive to minimum ionizing backgrounds that normally plague WIMP searches but retain sensitivity to the nuclear recoils that would be characteristic of WIMP scattering. 
 In a superheated liquid the process of radiation-induced bubble nucleation 
is described by the classical ``hot spike'' model~\cite{seitztheory}. For the phase transition to occur, the energy deposited by 
the particle must create a critically sized bubble, requiring a minimum energy deposition in
a volume smaller than the critical bubble. 
Under mildly superheated conditions, the latter requirement renders the bubble chamber insensitive to minimum ionizing particles. 

The radius of the critical bubble is given by the condition that the bubble be in (unstable) equilibrium with the surrounding superheated fluid~\cite{gibbs}. This demands the pressure balance
\begin{equation}
P_b - P_l = \frac{2\sigma}{r_c},
\end{equation}
where $P_b$ is the pressure inside the bubble, $P_l$ is the pressure in the liquid, $\sigma$ is the bubble surface tension, and $r_c$ is the critical bubble radius.  The pressure $P_b$ is fixed by the condition that the chemical potential inside and outside the bubble be equal, giving
\begin{equation}
(P_b-P_l) \approx (P_{sat}-P_l)\frac{\rho_l-\rho_v}{\rho_l},
\end{equation}
where $P_{sat}$ is the pressure in a saturated system at the given temperature, and $\rho_l$ and $\rho_v$ are the liquid and vapor densities in the saturated system~\cite{Peyrou}.

In Seitz's ``hot spike'' model for bubble nucleation, the entire energy necessary to create the critical bubble must come from the particle interaction that nucleates the bubble.  This is in contrast to earlier models that required only the work (free energy) to come from the particle interaction, with the remaining bubble-formation energy supplied by heat flowing in from the surrounding superheated fluid~\cite{glaser}.
As the name ``hot spike'' implies, the nucleation site in Seitz's model begins as a high-temperature seed, so it cannot draw heat from the surrounding fluid.

Once the decision is made to consider the total bubble creation energy rather than just the free energy, the threshold energy calculation is completely described by Gibbs~\cite{gibbs}.  This energy is given by
\begin{equation}
\begin{aligned}
E_T = \,&4\pi r_c^2 \left( \sigma - T\frac{\partial\sigma}{\partial{}T}\right) + \frac{4\pi}{3}r_c^3\rho_b\left(h_b - h_l\right) 
 - \\ & \frac{4\pi}{3}r_c^3\left(P_b-P_l\right) +  \mathcal{O}\!\left(\frac{\delta}{r_c}\right).
\end{aligned}
\label{eq:seitz}
\end{equation}
Here, $T$ is the temperature of the system, $\rho_b$ is the bubble vapor density, and $h_b$ and $h_l$ are the specific enthalpies of the bubble vapor and superheated liquid.  The surface tension $\sigma$ and temperature derivative are taken along the usual saturation curve.  The three terms give, from left to right, the heat necessary to create the bubble surface, the heat needed to vaporize the fluid to make the bubble interior, and a reversible work term done in expanding the bubble to the critical size that must be subtracted to avoid double-counting work present in both of the first two terms.  To good approximation $h_b-h_l$ may be replaced by the normal heat of vaporization at temperature $T$.

The greatest uncertainty in determining the thermodynamic $E_T$ is the relation between the surface tension at a flat liquid-vapor interface and the surface tension for a very small bubble.  This relation is described by the ``Tolman length''~$\delta$, which is unknown but is expected to be some fraction of the intermolecular distance~\cite{tolman}.  This translates to an uncertainty on $E_T$ of $\sim$$3\%$.  For the rest of this paper, we refer to the calculated threshold in Eq.~(\ref{eq:seitz}) as the Seitz threshold. 

The Seitz model assumes the efficiency for bubble nucleation is $100\%$ for all interactions that deposit $E \geq E_T$
over a volume small compared to the critical bubble. The length scales for nuclear recoil cascades in the energy region between 5 and 20 keV relevant for a WIMP search are similar to the critical radius, so the Seitz model may or may not give a good description of bubble nucleation, and direct calibrations of bubble nucleation efficiency are necessary.

   The working fluid discussed in this paper is iodotrifluoromethane or CF$_3$I, which contains two
highly sensitive WIMP target nuclei: fluorine, for spin-dependent interactions, 
and iodine, for spin-independent interactions.  Neutrons are typically used to mimic WIMPs in calibrating the nuclear recoil response of a WIMP detector, and neutron sources have been used to measure the nucleation threshold for carbon and fluorine recoils in CF$_3$I, CF$_3$Br~\cite{COUPPscience, COUPPnim} and C$_4$F$_{10}$~\cite{picassoCal} under various superheat conditions. However, iodine recoils contribute only a small fraction to the total neutron-nucleated 
bubble rate in CF$_3$I. Therefore, neutron sources are an ineffective calibration tool for iodine recoils in COUPP. We have used heavy daughter nuclei produced in alpha decays as a proxy~\cite{COUPPprd}, but these are high energy recoils of $\sim$100~keV. This paper describes a measurement of bubble nucleation efficiency for iodine recoils near our dark matter search thresholds.

\section{Experimental design}
    Our bubble chambers are insensitive to minimum ionizing 
particles, allowing us to exploit a new calibration technique using charged pions as WIMP surrogates to produce 
nuclear recoils by strong elastic scattering.  We measure the pion scattering 
angle using silicon pixel detectors. The 
nuclear recoil kinetic energy can be calculated by $E = (p\theta)^{2}/2M$ on an event by event basis, where $p$ is the 
beam momentum, $\theta$ the scattering angle, and $M$ the nuclear mass of the target. 

For a CF$_3$I target, a measured scattering angle corresponds to a different recoil energy depending on which nucleus is involved in the interaction; in this paper, we will refer to iodine equivalent recoil energy, $E_{Ie}$, as the energy given to an iodine nucleus for a given pion scattering angle. For a $\Pbeam$ pion beam, approximately $75\%$ of the rate of pions scattering into angles corresponding to $E_{Ie}$ between 5 and 20 keV is due to elastic scattering on iodine, with smaller contributions from carbon, fluorine, and inelastic scattering~\cite{BliedenPRD}. Therefore, the bubble nucleation efficiency for iodine recoils in a bubble chamber with Seitz threshold between 5 and 20 keV can be inferred from a measurement of the fraction of pion-scattering events that nucleate bubbles in the chamber as a function of $E_{Ie}$. 

    The measurement was performed in the Fermilab Test Beam Facility~\cite{FTBF} 
using a $\Pbeam$ mainly $\pi^-$ beam with $\sigma_p/p = 4\%$ and an angular spread of $<1.1$~mrad.  The absolute momentum of the beam is known to $3\%$. The pions were tracked with a silicon pixel 
telescope~\cite{Telescope} consisting of 4 upstream and 4 downstream silicon pixel plaquettes, with a spatial coverage of 14~mm x 14~mm. The total length of the telescope was 90~cm.   The angular resolution was 0.6~mrad~($\sigma$) in the horizontal ($x$) direction and 0.7~mrad in the vertical ($y$) direction, with roughly equal contributions from multiple Coulomb scattering (MCS) in the target and the spatial resolution of the telescope. Plastic scintillators triggered the pixel telescope on each beam particle.

  A small bubble chamber was designed for this measurement consisting of a quartz test tube with inner diameter 10~mm and 1-mm-thick wall, filled with 7~cm$^3$ of CF$_3$I.  The
small size is required to minimize MCS in the 
short radiation length of CF$_3$I ($58$~mm).   The bubble chamber was operated at a pressure of $30.0\pm0.1$~psia and a temperature of $34.2\pm0.2$~C with a nominal Seitz threshold of $\Tseitz$. The 
iodine equivalent threshold scattering angle is 4.7~mrad. An acoustic transducer was attached to the top of the test tube to record the acoustic emission produced during bubble formation, providing the time of bubble nucleation with $\sim$10~$\mu$s resolution.  Temperature control was provided by a water bath around the bubble chamber.

\begin{figure}[htbp]
\includegraphics[width=0.9\columnwidth,trim=0 0 0 0,clip=true]{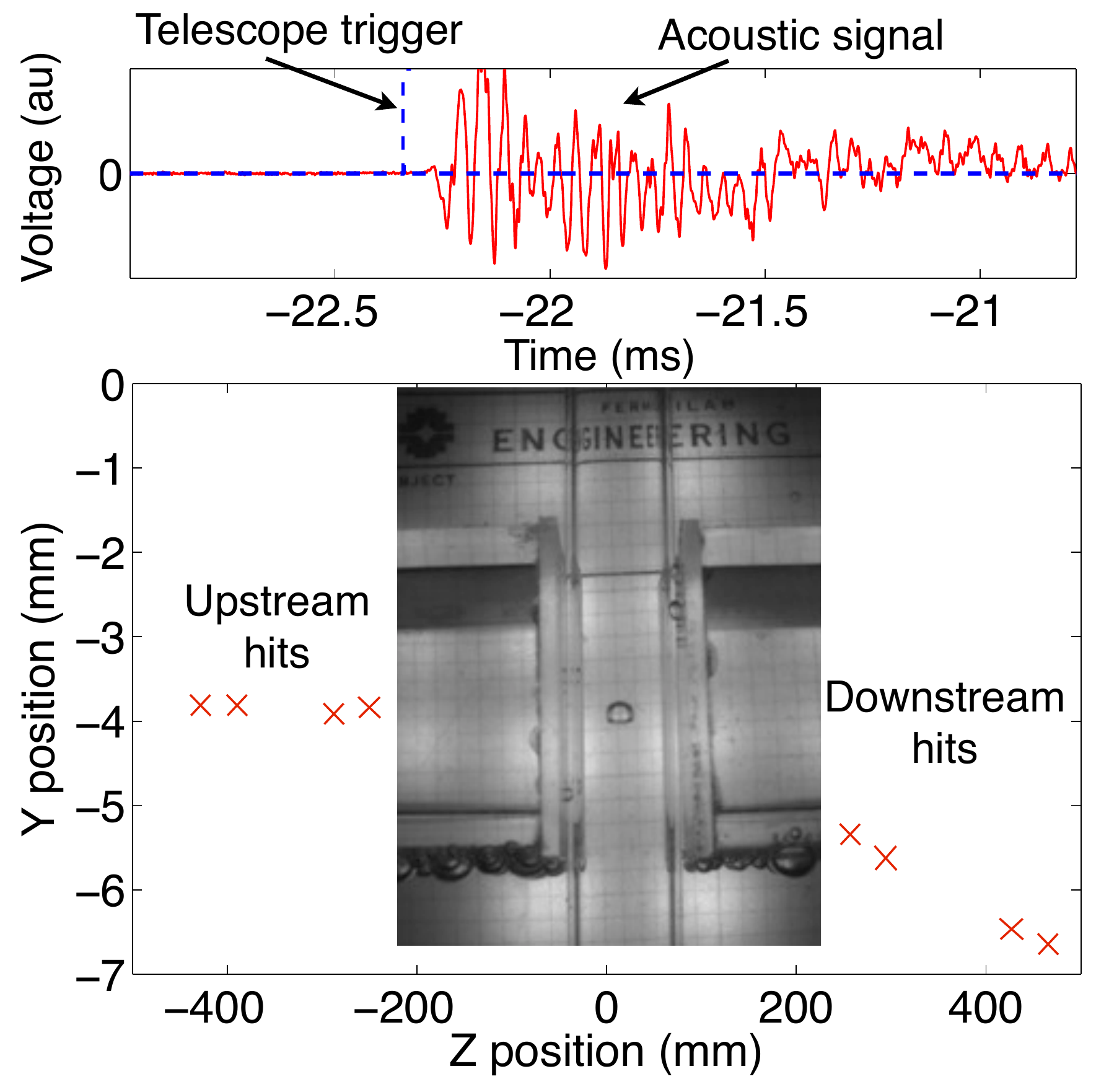}
\caption{\label{fig:signals} (Color online)
An example event ($\theta = 6$~mrad), including the relative timing of the telescope trigger and acoustic signal, one camera image of the bubble, and the $y$ and $z$ positions of the telescope hits. The pion beam is in the $+z$ direction. The camera image is not to scale but the test tube has inner diameter of 10~mm. Beam tubes in the water bath to minimize the material traversed by the pion beam are visible to either side of the bubble chamber.} 
\end{figure}

Bubble chamber data were taken between March 14 and March 28, 2012, with a 
beam flux of $\sim$1000 particles per 4-second beam spill with one 
spill per minute. The size of the beam spot was wider than both the bubble chamber and the pixel telescope.  The chamber was expanded to the superheated state 22 seconds before the arrival of the beam, allowing time for pressure and temperature transients to dissipate after expansion. The observation of bubbles by a 100-Hz video camera system created a bubble chamber trigger, causing the video images and associated data to be 
recorded and the chamber to be recompressed. After recompression, the chamber was dead for the remainder of the beam spill, allowing us to collect at most one bubble event per minute.  We collected about four good single-bubble events per hour, with the primary losses due to premature bubble chamber triggers, bubbles forming outside of the region covered by the telescope planes, multiple bubble events and large-angle scatters outside the acceptance of the downstream plaquettes. The last two categories are predominantly the result of inelastic interactions. Figure~\ref{fig:signals} shows an example scattering event. 

At the end 
of the run the CF$_3$I was removed and a target empty data set was taken.  In addition, data were taken in a test run in December 2011 with no target, as well as solid targets 
of quartz, graphite, Teflon or (C$_2$F$_4$)$_{\rm{n}}$, and crystalline iodine.

\section{Analysis} 
The primary analysis output is the bubble nucleation fraction as a function of $E_{Ie}$, given by the ratio
\begin{equation}
r_{\rm{bub}}=\frac{N_{\rm{bub}}}{(N_{\rm{tot}}-N_{\rm{multi}})f_{\rm{CF3I}}},
\label{eqn:ratio}
\end{equation}    
where $N_{\rm{bub}}$ is the observed number of pion tracks creating single bubbles, $N_{\rm{tot}}$ is the total number of pion tracks, $N_{\rm{multi}}$ is the number of tracks creating multiple bubbles, and $f_{\rm{CF3I}}$ is the fraction of scatters that occur in the active CF$_{3}$I volume, determined by a comparison of the number of scatters in the target-full data set to the number  in the target-empty data set normalized to the number of pion tracks ($N_{\rm{emp}}$): 
\begin{equation}
f_{\rm{CF3I}}=(N_{\rm{tot}}-N_{\rm{emp}})/N_{\rm{tot}}.
\end{equation} 
 An angular smearing correction is made 
to $N_{\rm{emp}}$ to include the MCS from the absent CF$_3$I by convolution with the standard Gaussian approximation for MCS~\cite{PDG}.

Each pion track is fitted for an upstream and downstream component, with an associated scattering angle and 3-D point of closest approach of the two components.  The upstream and downstream track segments are required to have exactly one hit cluster in at least three of 
the four pixel planes, good fits to straight lines ($\chi^2/\nu <4$), and to meet in space to within $0.5$~mm. 
 To exclude pions that passed through little or no CF$_3$I, the upstream track is required to pass within $4$~mm of the center of the $10$-mm-diameter bubble chamber in the $x$ direction.  The $y$ location of the track is limited by the vertical extent of the pixel planes.
Because the uncertainty on the location of the point of closest approach in the beam direction ($z$) depends strongly on the scattering angle, we require the $z$ location to be within 3$\sigma_z$ of the bubble chamber, where $\sigma_z$ is the uncertainty on $z$ for each individual event.
Events with more than one track are rejected.  As these track cuts are applied without regard to the presence of nucleations in the bubble chamber, their efficiency applies equally to $N_{\rm{bub}}$, $N_{\rm{tot}}$, $N_{\rm{multi}}$, and $N_{\rm{emp}}$, and therefore cancels in the final ratio, $r_{\rm{bub}}$.  

\begin{figure}
\includegraphics[width=0.9\columnwidth,trim=80 200 80 200,clip=true]{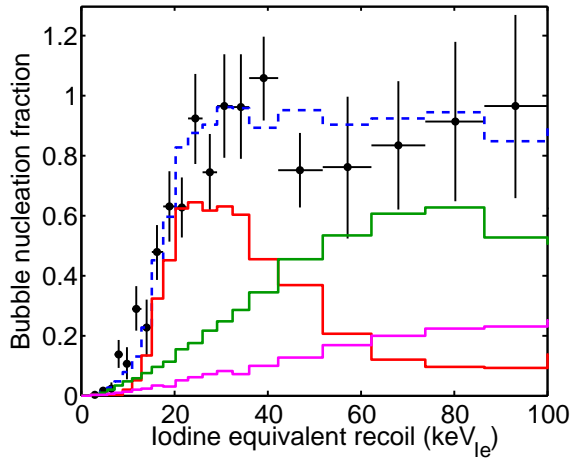}
\caption{\label{results} (Color online)
 The fraction of pion scattering events that produced bubbles as a function of iodine equivalent recoil energy.  The solid curves show the simulated contribution from individual recoil species (from high to low at 20 keV, red for iodine, green for fluorine, and pink for carbon and inelastics), with the blue dashed curve showing the sum. The iodine curve shown takes a step function efficiency model for iodine recoils using the best fit threshold of $\Tcirte$.}
\end{figure}

The next step is to associate a bubble with a unique track using both time and space correlations. The timing requirement for correlating a track with a bubble is chosen to be $20 < \Delta t < 120$~$\mu$s. The bubble locations are reconstructed using standard COUPP techniques~\cite{COUPPscience}, and the difference between reconstructed bubble position and point of closest approach of the track components is required to be less than 2.1~mm in the $x$ direction and less than 0.9~mm in the $y$ direction. The combined event acceptance of these timing and spatial cuts is $\eff$.  After these data selection and quality cuts, $\Nevt$ good single bubble events remain.  The final bubble nucleation fraction, $r_\mathrm{bub}$, 
is shown as the points in Fig.~\ref{results}.

    To disentangle the iodine component from carbon, fluorine and inelastic scattering, we perform a full simulation using GEANT4.9.5~\cite{geant}. The simulation was validated by comparing the simulated scattering angular distributions to data for no target, target empty, target full, and the solid targets. In all cases, in the MCS-dominated small scattering angle region there is good (few percent) agreement with no
adjustable parameters, suggesting that the telescope geometry is accurately modeled in the simulation.   In the 
larger scattering angle region dominated by strong elastic scattering, the 
simulation systematically overestimates the observed scattering rate by 
$\sim40\%$.  As this ratio is measured to be the same for Teflon ($1.45\pm0.10$) 
and  iodine ($1.41\pm0.12$), we assume that the relative contributions of iodine, fluorine and carbon are being accurately described by the MC.

A significant systematic uncertainty is introduced by our developing understanding of the carbon and fluorine recoil nucleation efficiency in this low energy regime. Ongoing studies with ad hoc neutron sources~\cite{JuanYBe} will reduce this uncertainty in the future, but here we apply the exponential carbon and fluorine efficiency model from~\cite{COUPPprd}:

\begin{equation}
\epsilon(E) = 1 - \exp(-\alpha[(E - E_T)/E_T]),
\label{eq:expo}
\end{equation}
where $E$ is the nuclear recoil energy and $E_T$ is the threshold energy. In ~\cite{COUPPprd} we found $\alpha=0.15$ to be consistent with AmBe neutron calibration data. 

We test the hypothesis that the iodine recoil nucleation efficiency follows the nominal Seitz model of a step function ($\alpha_I \gg 1$) with 100\% efficiency above the Seitz threshold by fitting a step function to the data in the region $9<E_{Ie}<42$~keV, allowing $E_T$ to float. The fit returns $E_T = (16.8\,^{+0.8}_{-1.1})~\mathrm{keV}$, where the error bars are statistical.  This value is 2.1$\sigma$ higher than the Seitz model threshold $\Tseitz \pm6\%\mathrm{(sys)}$, where the systematic error includes absolute energy scale uncertainties of 3\% in the beam momentum and 1\% in the scattering angle stemming from uncertainty in the $z$ positions of the plaquettes.  The fit is shown as the dashed blue line in Fig.~\ref{results}. 

\begin{figure}
\includegraphics[width=0.9\columnwidth,trim=80 200 80 200,clip=true]{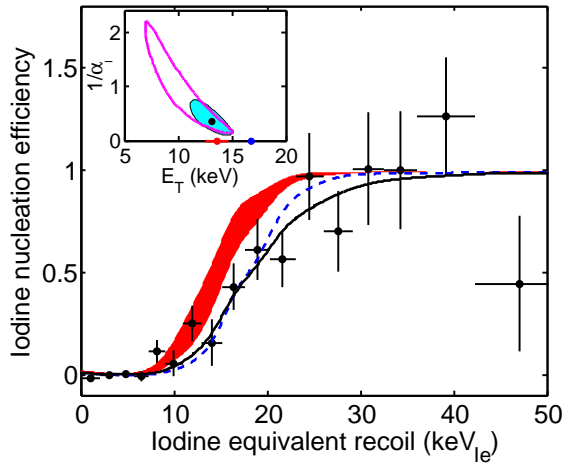}
\caption{\label{contour} (Color online)
 The data points represent the measured iodine nucleation efficiency as a function of iodine equivalent recoil energy, where the contributions from carbon, fluorine and inelastic scatters have been subtracted. The gradual turn on is predominantly due to the angular resolution of the experiment, as illustrated by both the red region, representing the step function model with the threshold varied within the uncertainty on the Seitz theory prediction, $E_T = (13.6\pm0.6)\, \mathrm{keV} \pm 6\%$, and the dashed blue curve, representing the best fit step function with $E_T = 16.8$ keV. The black curve shows the best fit exponential model with the threshold constrained by the theory as described in the text. The inset shows 2$\sigma$ contours for a fit to the exponential model with the threshold allowed to float (pink) or constrained by the theory (solid cyan). The colored dots represent the corresponding curves in the main plot.   }
\end{figure}

Figure~\ref{contour} shows the inferred iodine nucleation efficiency as a function of $E_{Ie}$ with the iodine component isolated by subtracting the simulated contributions from carbon, fluorine and inelastic scatters. The dashed blue curve is the best fit step function with $E_T = (16.8\,^{+0.8}_{-1.1})~\mathrm{keV}$. For comparison, the red region represents a step function at the predicted Seitz threshold, where the range represents the 1$\sigma$ band including the thermodynamic uncertainty and the scale uncertainties in the absolute energy scale of the experiment. 

Given the energy resolution smearing induced by MCS in this experiment, the preference for a value of $E_T$ higher than the prediction cannot be easily distinguished from an exponential model like Eq.~(\ref{eq:expo}) for iodine nucleation efficiency with a lower threshold energy and a finite value of $\alpha_I$. Previous studies have shown that the Seitz model accurately predicts the threshold at which bubble nucleation begins for heavy radon daughter nuclei in CF$_3$I~\cite{COUPPscience, COUPPprd}. We therefore perform a second fit applying the exponential model to iodine recoils, taking the Seitz threshold calculation as an external input to the analysis to explore the allowed range of $\alpha_I$. The best fit is shown as the black curve in Fig.~\ref{contour}. The inset shows 2$\sigma$ contours for fits to the exponential model with the threshold constrained by our prediction (shaded region) and free (unshaded region), along with the best fit points. 

 To assess the systematic errors associated with carbon and fluorine recoils, we refit the data with two alternative models for carbon and fluorine efficiency: the flat model from~\cite{COUPPprd} with energy-independent nucleation efficiency, $\eta_{C,F}=49\%$, above threshold, and a step function with $\eta_{C,F}=100\%$. The latter case represents the worst possible scenario for the response of the bubble chamber to iodine recoils, as the response to carbon and fluorine is maximized. We use the exponential model for iodine recoils, allowing $\alpha_I$ to float and treating $E_T$ as a nuisance parameter constrained by the prediction. The results of these fits are summarized in Table~\ref{tab:summary_alpha}. Extended fits over the energy interval $ 9 < E_{Ie} < 100$~keV have a negligible effect on the iodine fit parameters but disfavor the flat C/F efficiency model with $\eta_{C,F}=49\%$. 

\begin{table}[htb!]
\begin{tabular}{|l|c|c|c|}
\hline
C/F efficiency model& $\alpha_I$ & 90\% L.L & $\chi^2/\nu$ \\\hline
Flat model ($\eta_{C,F}=49\%$)& $4.5^{+4.4}_{-1.3}$ & $>2.8$ & 14.4/10 \\ 
Exp. ($\alpha_{C,F}=0.15$)& $2.8^{+1.6}_{-0.8}$ & $>1.8$ & 8.2/10 \\
Step function ($\eta_{C,F}=100\%$)& $2.2^{+0.9}_{-0.7}$ & $>1.4$ & 6.6/10 \\  \hline
\end{tabular}
\caption{Summary of fits to $\alpha_I$, including $90\%$ lower limits on $\alpha_I$. The three different C/F efficiency models described in the text are tested, and in all cases the predicted Seitz threshold is treated as a nuisance parameter. By maximizing the subtracted C/F contribution, the step function with $\eta_{C,F}=100\%$ represents the worst case for iodine efficiency.}
\label{tab:summary_alpha}
\end{table}

In conclusion, we have directly measured the efficiency for iodine bubble 
nucleation in a CF$_3$I bubble chamber operated with a nominal threshold of 
$\Tseitz$. For some models of carbon and fluorine efficiency, the response to iodine recoils is consistent with a step function at the Seitz threshold, but in all cases there is a preference for either a softer turn on or a slightly higher threshold. Even in the worst case scenario for iodine, however, the response of the chamber to iodine recoils is much closer to the nominal Seitz model than it is for carbon and fluorine recoils. This was expected from the considerably larger stopping power of iodine, which facilitates the concentration of energy that leads to critical bubble formation. Systematic uncertainties from both the absolute beam momentum calibration and the carbon and fluorine response limit the present measurement.  This measurement provides confirmation of the sensitivity of COUPP bubble chambers to spin-independent WIMP interactions with iodine nuclei, a confirmation that was not attainable using standard neutron source calibrations. The technique of employing hadron elastic scattering as a tool to measure bubble nucleation thresholds is 
now established, enabling the measurement of bubble nucleation energies on an event by event basis. 
We have begun studies of the feasibility to repeat this 
technique with different fluids.

The COUPP collaboration would like to thank Fermi National Accelerator 
Laboratory, the Department of Energy and the National Science Foundation for 
their support including grants PHY-0856273, PHY-1205987, PHY-0937500 and 
PHY-0919526.  We acknowledge technical assistance from Fermilab's Accelerator, Computing, 
and Particle Physics Divisions, and from A. Behnke at IUSB.



\begin{thebibliography}{99}



\bibitem{COUPPscience}
E.~Behnke {\it et al.},
Science\ {\bf 319}, 933 (2008).

\bibitem{picasso}
S.~Archambault {\it et al.},
Phys. Lett. {\bf B} 711, 153 (2012).

\bibitem{SIMPLE}
M.~Felizardo {\it et al.},
Phys. Rev. Lett. 108, 201302 (2012).


\bibitem{seitztheory}
F.~Seitz, Phys. Fluids {\bf 1}, 2 (1958).


\bibitem{gibbs}
Gibbs, On the Equilibrium of Heterogeneous Substances in {\bf The Scientific Papers of J. Willard Gibbs}, Ox Bow Press, 1993.

\bibitem{Peyrou}
 Peyrou, in {\bf Bubble and Spark Chambers, Vol I},Academic Press, 1967.
 
\bibitem{glaser}
 D.~A.~Glaser, Nuovo Cimento {\bf 11}, Suppl 2, 361 (1954).

\bibitem{tolman}
R.~Tolman, J. Chem. Phys. {\bf 17}, 333 (1949).


\bibitem{COUPPnim}
W.~J.~Bolte {\it et al.}, Nucl. Inst. Meth. A, {\bf 577}, 569 (2007).

\bibitem{picassoCal}
S.~Archambault {\it et al.},
New J. Phys. {\bf 13}, 043006 (2011).

\bibitem{COUPPprd}
E.~Behnke {\it et al.}, Phys. Rev. D {\bf 86}, 052001 (2012).

\bibitem{BliedenPRD}
H.~R.~Blieden {\it et al.}, Phys. Rev. D {\bf 11}, 14 (1975).

\bibitem{FTBF}
\url{http://www-ppd.fnal.gov/FTBF/}

\bibitem{Telescope}
\url{http://www-ese.fnal.gov/T992/}

\bibitem{PDG}
J.~Beringer {\it et al.} (PDG), Phys. Rev.  D {\bf 86}, 010001 (2012).

\bibitem{geant}
S.~Agostinelli {\it et al.}, Nucl. Inst. Meth. A, {\bf 506}, 250 (2003).

\bibitem{JuanYBe}
J.~I.~Collar, arXiv:1303.2686v2 (2013).


\end{thebibliography}
\end{document}